\documentclass[aps,superscriptaddress,showpacs,nofootinbib]{revtex4}
\usepackage{graphics}
\usepackage{bm}

\input{epsf}

\newcommand{\be}{\begin{equation}}
\newcommand{\ee}{\end{equation}}
\newcommand{\bea}{\begin{eqnarray}}
\newcommand{\eea}{\end{eqnarray}}

\begin{document}

\preprint{UG-FT-211/06}

\preprint{CAFPE-81/06}

\title{Warm inflation dynamics in the low temperature regime}
\author{Mar Bastero-Gil}
\email{mbg@ugr.es}
\affiliation{Departamento de F\'{\i}sica Te\'orica y del Cosmos,
  Universidad de Granada, Granada-18071, Spain}

\author{Arjun Berera}
\email{ab@ph.ed.ac.uk}
\affiliation{ School of Physics, University of
Edinburgh, Edinburgh, EH9 3JZ, United Kingdom}

\begin{abstract}
Warm inflation scenarios are studied with the dissipative coefficient
computed in the equilibrium approximation. Use is made of the
analytical expressions available in the low temperature regime
with focus on 
the possibility of achieving strong dissipation within this
approximation. Two different types of models are examined: monomial
or equivalently chaotic type potentials, and hybrid like models where
the energy density during inflation is dominated by the false vacuum.
In both cases dissipation is shown to typically 
increase during inflation and bring the system into the strong
dissipative regime. Observational consequences are explored for the
amplitude of the primordial spectrum and the spectral index, which
translate into constraints on the number of fields mediating the
dissipative mechanism, and the number of light degrees of freedom
produced during inflation. 
This paper furthers the foundational
development of warm inflation dynamics from
first principles quantum field theory by calculating
conservative lower bound estimates on dissipative effects
during inflation using the well established thermal
equilibrium approximation.  This approximation does not
completely represent the actual physical system and earlier
work has shown relaxing both the equilibrium and
low temperature constraints can substantially enlarge the warm inflation
regime,  
but these improvements still need further
theoretical development.

\medskip                                   

\noindent
keywords: cosmology, inflation
\end{abstract}

\maketitle

\section{Introduction}
\label{sect1}
Cosmological observations are consistent with an early period of
inflation in the evolution of the Universe, 
which among other things accounts for the observed flatness of the
Universe and give rise to the primordial curvature perturbation.  
The quest now is for a realistic particle physics model of inflation,
and a better understanding of the inflationary dynamics. 
The model has also to include the mechanism for the subsequent
transition from an inflationary universe into a radiation dominated
one. This requires the inflaton to couple to other particles, 
to allow for the conversion of the vacuum energy into
radiation at the end of inflation during the reheating period. 
In addition, these interactions can
lead to dissipative effects already relevant during inflation,
affecting the inflationary predictions. These kind of scenarios are
called warm inflation \cite{warm,Moss:wn,Berera:1995wh}. 

In the warm inflationary models, the dissipative term appears as an
extra friction term in the evolution equation for the inflaton field $\phi$,
and the corresponding source term for radiation $\rho_R$:
\bea
\ddot \phi + 3 H \dot \phi + \Upsilon_\phi \dot \phi + V_\phi &=&0
\,,\label{eominf}\\
\dot \rho_R + 4 H \rho_R  &=& \Upsilon_\phi \dot \phi^2\,, \label{eomrad}
\eea
where $V_\phi$ denotes the derivative of the potential with respect to
the inflaton field.
Warm inflation scenarios exhibit several features that
are attractive for model building.  For one, dissipation
allows the inflaton mass during inflation
to be much bigger than the Hubble scale \cite{Berera:1999ws,bb1},
thus completely avoiding
the ``eta problem'' 
\cite{Copeland:1994vg,Dine,Gaillard:1995az,Kolda:1998kc,randall}, 
which is a generic problem in
standard cold inflation scenarios in supergravity (sugra) theories.
Another attractive model building feature is for monomial
potentials, inflation occurs with the inflaton amplitude
below the Planck scale, $m_{P}=2.4\times 10^{18}$ GeV
\cite{Berera:1999ws,bb1}. In contrast, for 
monomial potentials in the cold inflation case, usually
called chaotic inflation scenarios \cite{ci},
the inflaton amplitude during inflation is larger than the
Planck scale.  This is a problem for model building, since
in this case the infinite number of nonrenormalizable operator
corrections,
$\sim \sum_{n=1}^{\infty} g_n \phi^{4+n} (\phi/m_{P})^n$
would become important and so have to be retained \cite{randall}.

The challenge of realizing warm inflation is in understanding
the dissipative dynamics from first principles quantum field
theory.  In terms of analytic approximations,
much work has already been devoted to this
problem \cite{bgr,br,br05,Hall:2004zr,ian}.
Inevitably warm inflation dynamics is nonequilibrium and a complete
treatment of the statistical state during warm inflation is
probably not amenable to simple analytic approximations,
with more numerical based methods such as
\cite{Lawrie:2002wm,2pi} needed.
Nevertheless, progress must be made systematically, initially
understanding what can be from analytic and near analytic treatments.
There are two known analytic approximations that would be
relevant to apply to this problem. First is a quasiparticle
approximation following Morikawa and Sasaki \cite{Morikawa:dz}.
The other is the equilibrium approximation, which is based on
the assumption that the statistical state of all fields
always remains in thermal equilibrium \cite{ian,boy}.
The quasiparticle approximation has been developed for warm inflation
in \cite{br,br05,Hall:2004zr} and
several interesting model building possibilities
have been found in \cite{bb1}.  The equilibrium approximation has
been developed for warm inflation in \cite{ian}.
In this paper we will explore some of the interesting
warm inflation models that such an approximation yields.

Testing the equilibrium approximation has an important general
significance to the overall understanding of dissipative effects
during inflation.
In this approximation, the basic assumption is that
the field system is minimally disturbed and so this approximation
likely provides a lower bound on the degree of dissipative
effects and particle production during inflation.
Thus models that work under this approximation
provide a minimal expectation on the overall
robustness of warm inflation.
Such solutions provide crucial existence proofs
of the viability and consistency of warm inflation models
with quantum field theory, which has been a basic question
about these scenarios since they were first suggested \cite{bgr,yl}.
This of course is of general significance in the development
of inflationary dynamics, since an alternative way to
state these results is regimes which would have been regarded
unquestionably to be governed by standard cold inflation
dynamics in fact are not, since dissipative effects are shown
to be significant.  For the better part of the existence of
the inflation idea, the blanket assumption has always been the dynamics
is cold inflationary.  The observation was made much
later in the development of the inflation subject  \cite{warm}
that in fact this is not
the unique situation and that dissipative effects can occur
during inflation.  Thus the results found in this paper provide
a significant step in breaking these preset early notions about
inflation dynamics and demonstrating that dissipation is
a generic feature during inflationary expansion.

In the high temperature regime, in earlier work
a class of warm inflation models
were found for the equilibrium approximation in \cite{bgr};
however one of the pronounced features of these models were the
requirement of a very large number of fields $\sim 10^4$,
and so not attractive for most model building purposes
(although see \cite{bk} for string motivated models).
In this paper, we will show that in the low-temperature
regime, many types of reasonable warm inflation models
can be found, in particular for a moderate number of fields
and sensible values of the parameters.

The specific field theory models examined in this paper
are the same as in \cite{br,br05}, in which the inflaton field
is coupled to a heavy bosonic field which in turn is
coupled to light fields.
An important feature of such coupling configurations
is that even for large perturbative couplings,
in supersymmetric (susy) models, the quantum corrections to the effective
potential from these terms can be controlled enough
to maintain an adequately flat inflaton 
potential \cite{br05,Hall:2004zr},
yet susy provides no cancellation of the time nonlocal terms,
which are the dissipative terms, and so such terms can
be quite significant.

The basic picture of dissipative dynamics for this class of models
is as the background inflaton field moves down the
potential, it excites the heavy bosonic fields
which in turn decays into light degrees of
freedom \cite{br,br05,Hall:2004zr}. The latter quickly thermalize and become 
radiation. Consistency of the approximations then demands the
thermalization timescale to be smaller than the evolution timescale of
the inflation field and the expansion timescale of the Universe. In
addition, the condition $T \gg H$ allows  to neglect the expansion of
the universe  when computing $\Upsilon_\phi$. Under these conditions,
the dissipative coefficient $\Upsilon_\phi$ has been
recently computed for a generic suspersymmetric inflationary model, in a
thermal approximation \cite{ian}. It was shown that in the low $T$
limit, with $H \ll T < m_\chi$, the dissipative coefficient behaves as
$T^3/m_\chi^2$, $m_\chi$ being the mass of the mediating field. Still,
dissipation can be large enough to dominate over the expansion rate
and bring the inflaton into the strong dissipative regime with
$\Upsilon_\phi > H$.    

In this paper we shall explore which kind of models of inflation can be
brought into the strong dissipative regime within the low $T$
approximation.   
We want to explore the viability of the scenario by
checking whether we can satisfy the conditions $\Upsilon_\phi > H$ and
$H < T < m_\chi$ for at least 50-60 e-folds. Due to the
$T$ dependence on  
the dissipative coefficient, the ratio $T/m_\chi$ tends to increase
during inflation, and at some point the low $T$
approximation breaks down. The system would  move then into the high
$T$ regime, where the dissipative coefficient goes linearly with $T$,
and soon after that $\Upsilon_\phi$ drops below $H$ and inflation
ends. The transition from the low to the high $T$ regime may last some
e-folds, but taking properly into account this period would require a
more involved calculation of the dissipative coefficient beyond the
analytical approximation. Therefore, as a first step in studying this
kind of models, we only work within the low $T$ approximation,
assuming that inflation ends soon after the condition $T < m_\chi$ is
violated.  

In section \ref{sect2} we briefly review the basics of the warm
inflationary dynamics in the strong dissipative regime, and applied
this to some generic inflationary models, divided into two groups:
(a) monomial potentials or chaotic inflation models, and (b) small field
models where a constant term dominates the potential energy. In
section \ref{sect3} we summarize our findings and further comment on
the consequences for model building. 

\section{Dissipation in the low $T$ regime}
\label{sect2}

We consider the interactions given in the superpotential:   
\be
W= g \Phi X^2 + h X Y^2  \,,
\ee
where $\Phi$, $X$, and $Y$ denote superfields, and $\phi$, $\chi$, and
$y$ will refer to their bosonic components. 
During inflation, the field $y$ and its fermionic partner $\tilde y$ remain massless, while the mediating field $\chi$ gets its mass from the
interaction with the inflaton field $\phi$, with $m_\chi= \sqrt{2} g \phi$. 
Following Ref. \cite{ian}, the
dissipative coefficient in the low $T$ 
regime is well approximated by:
\be
 \Upsilon_\phi \simeq 0.64 \times g^2 h^4 \left(\frac{g \phi}{m_\chi}\right)^4
\frac{T^3}{m_\chi^2} \,.   
\label{upsilon0}
\ee
Alternatively, we may consider a susy hybrid model of inflation,
with the mass of the $\chi$ field given by $m_\chi^2= 2 g^2 (\phi^2 -
\phi_c^2)$, with $\phi_c$ being the critical value. In either case,  
 the dissipative coefficient given in Eq. (\ref{upsilon0}) does not
depend on the coupling $g$ in this regime, but only on  
$h$ which can be quite large, $h \simeq O(1)$. The numerical
coefficient in Eq. (\ref{upsilon0}) was computed taking $X$, $Y$ to be
singlet complex fields. But in principle, they may belong to larger
representations  of a Grand Unification Theory (GUT) group, as it is
typically assumed in susy 
hybrid models. This will give rise to an extra factor of ${\cal N}=N_\chi
N_{decay}^2$ in front of the dissipative coefficient, where
$N_\chi$ is the multiplicity of the $X$ superfield, and $N_{decay}$
counts the no. of decay channels available in $X$'s 
decays.  Taking $m_\chi^2 = 2 g^2 \phi^2$, we have then: 
\be
 \Upsilon_\phi \simeq C_\phi \frac{T^3}{\phi^2} \,,
\label{upsilon}
\ee
where $C_\phi =0.16\times h^4 {\cal N}$.

The conditions for slow-roll
inflation are modified due to the extra 
friction term in Eq. (\ref{eominf}). Demanding $\ddot \phi < (3 H + \Upsilon_\phi) \dot \phi$ and $\dot \phi^2 < V$, they are given now by:
\bea
\eta = \frac{\eta_H}{(1+r)^2} &<& 1 \,, \label{eta}\\
\epsilon = \frac{\epsilon_H}{(1+r)^2} &<& 1 \label{epsilon} \,, \\
\frac{V_\phi/\phi}{3 H^2} \frac{r}{(1 + r)^3} &<& 1 \,, 
\eea
where $\eta_H= m_P^2 (V_{\phi \phi})/V$, $\epsilon_H=m_P^2
(V_\phi/V)^2/2$ are the standard slow-roll parameters, and
$r=\Upsilon_\phi/(3H)$. In addition, once the source term for the
radiation dominates in Eq. (\ref{eomrad}), this would reduce to:
\be
4 H \rho_R \simeq \Upsilon_\phi \dot \phi^2 \,. \label{radsr}
\ee  
However, in order to have  $\dot \rho_R < 4 H \rho_R$ and then
Eq. (\ref{radsr}), we need to impose instead the slow-roll conditions: 
\bea
\frac{\eta_H}{(1+r)} &<& 1 \,, \\
\frac{\epsilon_H}{(1+r)} &<& 1 \,, \\
\frac{V_\phi/\phi}{3 H^2 (1 + r) } &<& 1 \,. 
\eea
For the kind of potentials considered in the following, we would have
$\epsilon_H \ll (V_\phi/phi)/(3 H^2) \simeq \eta_H$, and therefore the
slow-roll conditions reduce mainly to:    
\be 
\frac{V_\phi/\phi}{3 H^2 (1 + r) } \simeq \frac{\eta_H}{(1+r)} < 1
\,. 
\ee
Once this is fulfilled, also the conditions  $\eta <1$, $\epsilon <1$
are satisfied. The above 
condition ensures that the radiation does not increase too
fast during warm inflation, otherwise it will dominate
too soon, not allowing inflation too proceed. 

Therefore, in the strong dissipative regime with $r > 1$, the
slow-roll evolution equations for the inflaton field and the radiation
are given by: 
\bea
\dot \phi &\simeq& -\frac{V_\phi}{\Upsilon_\phi} \,, \label{phidot}\\ 
\rho_R &\simeq& \frac{V_\phi^2}{ 4 H \Upsilon_\phi} \label{rhor}\,. 
\eea
Alternatively, using $\rho_R = C_R T^4$, we have for the temperature
of the thermal bath:
\be
T \simeq \left( \frac{(V_\phi \phi)^2}{4 H C_\phi C_R} \right)^{1/7}
\label{temp}\,, 
\ee
where $C_R= \pi^2 g_*/30$, and $g_*$ is the effective number of
light degrees of freedom. Once all the particles are in thermal
equilibrium at a common $T$ through rapid interactions, $g_*$ counts
the number of relativistic degrees of freedom in the model. However,
thermalization is not an instantaneous process, and for a particle
specie kinetic equilibrium with the thermal plasma is only reached
once its interaction rate becomes larger than the Hubble rate. Taking
into account the thermalization rate would translate into a lower
effective $T$ than expected \cite{thermalisation}, or equivalently a
lower value $g_*(T)$. Therefore, we may expect also $g_*(T)$ varying 
while in the dissipative regime, from say $O(10)$ to the 
value for the Minimal Supersymetric Standard Model (MSSM) $g_*= 228.75$ .

\subsection{Monomial potentials}
We first study the inflationary trajectory for a general inflaton
monomial potential:  
\be
V(\phi) = V_0 \left(\frac{\phi}{m_P}\right)^n \label{vpoly}\,,
\ee
with $n>0$. Without enough dissipation, i.e., either  for cold
inflation with $r=0$, or only weak dissipation  with $r < 1$, these
kind of models lead to inflation 
only for values of the inflaton field larger than the Planck mass $m_P$. On
the other hand, in the strong dissipative due to the larger friction
term slow-roll conditions  Eqs. (\ref{eta}) and  (\ref{epsilon})
can be fulfilled for values of the field well below Planck. That is,
we can regard now Eq. (\ref{vpoly}) from the  effective field theory
point of view, with the potential well define below the cut-off scale
$m_P$; higher order term contributions suppressed by $m_P$ will be then
negligible, without the need of fine-tuning the coefficients in front.  

During slow-roll, once the system enters in the strong dissipative regime, 
we can integrate Eq. (\ref{phidot}) using  Eqs. (\ref{temp}) and
(\ref{upsilon}). The potential decreases from its initial value $V(0)$ as:
\be
\frac{V}{V(0)}\simeq \left( 1- \frac{4 \rho_R(0)}{7 V(0)} N_e
\right)^7 \,, 
\label{Vchaotic}
\ee
where $N_e$ is the no. of e-folds from the beginning of inflation, an
$\rho_R(0)$ is the initial value for the radiation given by:
\be
\frac{\rho_R(0)}{V(0)}\simeq \left(\frac{9 n^8}{4} \frac{C_R^3}{C_\phi^4}
\left(\frac{m_P^4}{V(0)}\right)\right)^{1/7} \,,
\ee
Eq. (\ref{Vchaotic}) is
only valid in the strong dissipative regime, $r >1$ , with:
\be
r\simeq \left(\frac{ n^6 C_\phi^4}{576 C_R^3}\right)^{1/7} 
\left(\frac{m_P}{\phi(0)}\right)^2
\left(\frac{V(0)}{m_P^4}\right)^{1/7}
\left(\frac{V(0)}{V}\right)^{2/n-1/7} \,, \label{rchaotic}
\ee
which increases during inflation for $n < 14$. 
On the other hand, the values of $\eta_H/r$, $T/H$ and $T/\phi$ are
given respectively by:  
\bea
\frac{\eta_H}{r} &\simeq& 4 \left(\frac{n-1}{n}\right)\left(
\frac{\rho_R(0)}{ V(0)}\right)  
\left(\frac{V(0)}{V}\right)^{1/7} \,, \label{etaHrchaotic}\\
\frac{T}{H} &\simeq& \left(\frac{81 n^2}{4 C_\phi C_R} \right)^{1/7}
\left(\frac{m_P^4}{V(0)}\right)^{2/7}\left(\frac{V(0)}{V}\right)^{2/7}
\,, \label{THchaotic}\\  
\frac{T}{\phi} &\simeq& \left(\frac{m_P}{\sqrt{3}\phi(0)}\right)\left(
\frac{81 n^2}{4 C_\phi C_R}\right)^{1/7}
\left(\frac{V(0)}{m_P^4}\right)^{3/14}
\left(\frac{V}{V(0)}\right)^{3/14-1/n} \,. \label{phiTchaotic}
\eea
Therefore, all the above ratios increase as inflation proceeds. This
means that after some no. of e-folds, we will either have $T/\phi >
0.1$, and the  low $T$ approximation will not longer hold,  or
$\eta_H/r > 1$ , and  inflation will end shortly after when $\rho_R > V$. From
(\ref{etaHrchaotic}) and (\ref{THchaotic}) we have the relation: 
\be
 \frac{\eta_H}{r} \simeq 4 \left(\frac{n-1}{n}\right)\left(
 \frac{\rho_R}{ V}\right) \,,
\ee
and for any power $n > 1$ we have that the radiation would not
dominate the total energy density in the slow-roll regime with
$\eta_H/r < 1$. Once this ratio grows beyond one, so does the ratio
$\rho_R/V$, and inflation ends when the radiation becomes larger than
the inflaton potential. 

\begin{figure}[t] 
\hfil\scalebox{0.5} {\includegraphics{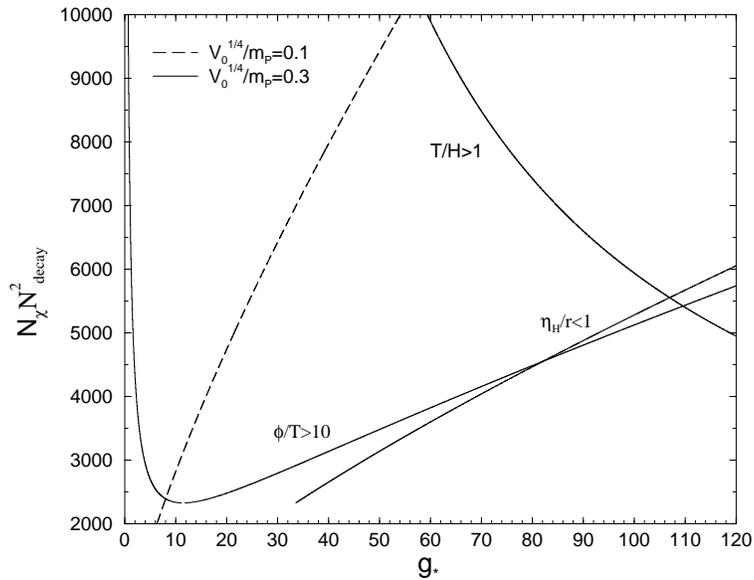}}\hfil
\caption{Quartic potential: Allowed values for $g_*$ 
  and ${\cal N} \equiv N_{\chi} N_{decay}^2$
  for having at least 50 e-folds of strong warm inflation: (a) enclosed
  region by the solid  lines  when $V(0)^{1/4}/m_P=0.3$; (b) to the
  left of the dashed line when $V(0)^{1/4}/m_P=0.1$. We have taken
  $\phi(0)/m_P=1$. }
\label{plot1}
\end{figure}

Without a specific model at hand,  we can always consider ${\cal N}=N_\chi
N^2_{decay}$ and $g_*$ 
as free parameters\footnote{We have also the value of the coupling $h$
among the heavy field $\chi$ and the light degrees of freedom, which
hereon we fix to its maximum allowed value, $h=\sqrt{4 \pi}$ when
giving bound on the value of ${\cal N}$.}, 
and  see for which values we can keep $\eta_H/r$
and $T/\phi$ small enough  for at least 50 e-folds or so (and $T/H
>1$). In turn, these values  will depend on  the initial value  for
$V(0)$, and in the case of $\phi/T$ also on $\phi(0)$. For example,
in Fig. (\ref{plot1})  we have plotted the allowed region in
the plane $g_*-{\cal N}$  for a quartic potential taking $V(0)^{1/4}
\simeq 0.3 m_P$ and $\phi(0)= m_P$; in this case, in order to satisfy all the
constraints we require $g_* < 100$ but ${\cal N} > 2300$. Similar
results are obtained for other powers of the potential. By lowering
the value of the potential, it is easier to fulfill all conditions
except that for the ratio $\eta_H/r$. Keeping the latter below one
gives the lower bound:
\be
{\cal N} > 8.4 \times 10^{-2} g_*^{3/4} \left( \frac{m_P}{V(0)^{1/4}}
\right) \left( n^{1/7} (n -1) + \frac{n}{7} N_e \right)^{7/4} \,,
\ee 
and the lower $V(0)$ is, the larger ${\cal N}$ has to be. For example,
for $n=4$, $V(0)^{1/4}/m_P = 0.1$, and $g_*= 10$ we would need ${\cal
  N} > 2800$ (dashed line in Fig. (\ref{plot1})), but getting to the
no. of degrees of freedom for the  
MSSM, $g_*= 228.75$, would require ${\cal N} > 29000$, which looks
rather large. In any case, curiously enough, due to the extra friction
we can have inflation for values of the field below Planck, but the
model prefers an initial value of the height of the potential only an
order of magnitude or so below Planck.  

On the other hand, the amplitude of the primordial spectrum is also
affected by the strong dissipative friction term and the presence of a
thermal bath. Approximately, when $T>H$ the fluctuations of the
inflaton field are induced by the thermal fluctuations, instead of
being vacuum fluctuations, with a spectrum proportional to the
temperature of the thermal bath. In
particular, when $r \gg 1$ we have for the spectrum of the inflaton
fluctuations \cite{Berera:1999ws}: 
\be
P_{\delta \phi}^{1/2} \simeq \left(\frac{\pi r}{4}\right)^{1/4}
\sqrt{T H} \,,
\ee
with the amplitude of the primordial spectrum of the curvature
perturbation given by: 
\be
P_{\cal R}^{1/2} \simeq \left | \frac{H}{\dot \phi}\right| P_{\delta
  \phi}^{1/2}  \simeq \left| \frac{3 H^3}{V_\phi}\right| \left(\frac{\pi
  r}{4}\right)^{1/4} (1 + r) \sqrt{\frac{T} {H}}\,, \label{pr}
\ee
which in the case of the monomial potential is given by: 
\be  
 P_{\cal R}^{1/2} \simeq \left( \frac{\pi}{12}\right)^{1/4}
\left( \frac{n^3}{6}\right)^{17/14}
\left( \frac{C_\phi^{9/14}}{C_R^{17/28}}\right)
\left( \frac{m_P}{\phi}\right)^{3/2}
\left( \frac{V}{m_P^4}\right)^{15/28} \,. 
\label{prchaotic}
\ee
In order to keep the amplitude of the primordial spectrum consistent
with WMAP's value \cite{WMAP}, $P_{\cal R}^{1/2} \simeq 5.5 \times
10^{-5}$, we 
would need a potential much smaller than $O(10^{-12} m^4_P)$. But for such
a value of the potential, we would need roughly ${\cal N} \sim
O(10^6)$ in order to get at least 50 e-folds in the strong dissipative
regime, which seems rather large and unnatural\footnote{Strictly
  speaking, the numbers
  and order of magnitude estimations have been obtained for the quartic
monomial, but similar values are obtained for the quadratic and
other powers.}. Therefore, although in
principle having strong dissipation during inflation is possible with
a monomial potential, maintaining inflation for at least 50 e-folds  
requires rather large values of ${\cal N}$, and it tends to produce a
too large amplitude for the primordial spectrum.

\subsection{Hybrid like models of inflation}
We consider now small field models of inflation, by adding a constant
term to the monomial potential:  
\bea
V(\phi) &=& V_0 \left( 1 + \left(\frac{\phi}{M}\right)^n \right)  
\label{vhybrid}\,,\;\;\; n > 0 \\
V(\phi) &=& V_0 \left( 1 + \beta \ln\frac{\phi}{M}\right) 
\label{vshybrid}\,,\;\;\; n = 0 
\eea
Given that during inflation the potential is
dominated by the constant term $V_0$, we can easily keep the value of the
field below Planck in this class of models during slow-roll
inflation. We regard this kind of 
potential as a generalization of a hybrid model \cite{linde,Copeland:1994vg},
where inflation ends  once the inflaton field reaches the critical value, 
destabilizing the waterfall field coupled to it. Those interactions are not
relevant to study the slow-roll dynamics, only to mark the end of
inflation, and therefore we do not need to consider them in the
inflationary potential Eqs. (\ref{vhybrid}), (\ref{vshybrid}). 
However, 
the same  interactions between the inflaton and the waterfall field
required by the hybrid mechanism will give rise to dissipation, and
leads to Eq. (\ref{upsilon}) in the low T regime. The case $n=2$ would
be the standard hybrid model \cite{linde,Copeland:1994vg}, with a mass term
for the inflaton, whereas when $n=0$ we have the susy model with the
logarithmic correction coming from the 1-loop effective potential 
\cite{dvali,lazarides}. In supersymmetric hybrid models, one needs to
worry about the so called ``eta'' problem 
\cite{Copeland:1994vg,Dine,Gaillard:1995az,Kolda:1998kc,randall}, i.e., the
fact that generically sugra corrections give rise to scalar masses of
the order of the Hubble parameter, including that of the inflaton,
then forbidding slow-roll inflaton. Different solution to this problem
exist in  the literature, for example by combining specific forms of
the superpotential and the K\"ahler potential
\cite{Copeland:1994vg,dvali}. Nevertheless, typically although we can avoid
the quadratic correction, i.e, a mass contribution,  sugra corrections
manifest as higher powers in the inflaton field \cite{sugra}. In
the case of strong  warm inflation, the 
presence of the extra friction term alleviates the problem: slow-roll
conditions are fulfilled also for inflaton masses in the range $ H <m_\phi <
\Upsilon_\phi$. In addition, the values of the field being smaller
than in standard cold inflation, the effect of higher order sugra
corrections is also suppressed.     

The Hubble rate remains practically constant and given by
$H \simeq V_0^{1/2}/(\sqrt{3} m_P)$ during inflation, so
neglecting its variation we can  integrate the
slow-roll equations in the strong dissipative regime,
Eq. (\ref{phidot}) together with  Eq. (\ref{temp}), and we obtain: 
\bea
\left(\frac{\phi}{\phi(0)}\right) &\simeq&   
\left( 1+ \frac{n}{7} \left(\frac{64 a^2 C_R^{3}}{C_\phi^4}\right)^{1/7}  
 \left(\frac{\phi (0)}{H}\right)^{n/7}
N_e \right)^{-7/n} \,, \label{phihybrid}\\
\left( \frac{T}{H} \right)^7&\simeq& \frac{a^4}{4 C_R
  C_\phi}\left(\frac{\phi}{H}\right)^{2 n} \,, \label{THhybrid}\\ 
\left(\frac{\phi}{T}\right) &\simeq&  \left(\frac{\phi(0)}{T(0)}\right) 
\left( \frac{\phi}{\phi(0)}\right)^{1-2n/7} \label{phiThybrid}\,, 
\eea
and therefore:
\bea
r &\simeq& \frac{C_\phi}{3} \left(\frac{a^4}{4 C_\phi C_R}\right)^{3/7} 
  \left(\frac{\phi}{H}\right)^{6n/7 -2} \,, \label{rhybrid} \\
\frac{\eta_H}{r} &\simeq& \left(\frac{\eta_H}{r} \right)_0
\left(\frac{\phi} {\phi(0)}\right)^{n/7} \,, \label{etahhybrid} \\
\frac{\rho_R}{V} &\simeq& \left(\frac{\rho_R}{V} \right)_0
\left(\frac{\phi} {\phi(0)}\right)^{2n/7} \,, \label{rhorV}
\eea
where the subscript ``0'' denotes the initial value and we have defined:
\be
a^2=\left\{\begin{array}{ll} \left(\frac{n
 V_0}{H^4}\right)\left(\frac{H}{M}\right)^n \,,&n\neq 0 \,,\\ 
\left(\frac{\beta V_0}{H^4}\right) \,,&n= 0 \,.\end{array}
\right.
\ee
The amplitude of the primordial spectrum is given by Eq. (\ref{pr}),
  and the prediction for the spectral index $n_S$ is given by:
\be
n_S-1 \simeq 2 \frac{d \ln P_{\cal R}^{1/2}}{d \ln k} \simeq 
2 \frac{\dot \phi}{H}\frac{d \ln P_{\cal R}^{1/2}}{d \phi} \simeq 
-\frac{2 V_\phi}{3 H^2 (1 +r)}\frac{d \ln P_{\cal R}^{1/2}}{d \phi} \,,
\ee
evaluated at $k_0$, corresponding to horizon exit ($k
= H a$) at say 50-60 e-folds before the end of inflation. Taking into account
  the $T$ field dependence, derived from $4 H \rho_R \simeq
  \Upsilon_\phi \dot \phi^2$, we get: 
\be
n_S -1\simeq \frac{1}{1+ 7 r} \left( -\frac{3(3 r -1)}{1+r} \epsilon_H - 3 \eta_H +
\frac{6(1 + 3 r)}{1 + r}\frac{m_P}{\phi} \sqrt{2 \epsilon_H} \right) 
  \,. \label{ns}
\ee
where $\eta_H$ and $\epsilon_H$ are the standard slow-roll parameters without
dissipation. 
By taking again the derivative of Eq. (\ref{ns}) with
respect to $\ln k$, i. e., with respect to the field, we obtain the
  running of the spectral index:  
\bea
n_S^\prime = \frac{d n_S}{d \ln k} &\simeq& \frac{1-n_S}{1+r} r^\prime
+\frac{1}{1+ 7 r} \left( -\frac{3 (3 r -1)}{1+r} \epsilon_H^\prime - 3 \eta_H^\prime +
\frac{6 (1 +3 r)}{ 1 + r} [\frac{m_P}{\phi} \sqrt{2 \epsilon_H}]^\prime \right) \nonumber \\
& & +\frac{1}{(1+r) ( 1 +7 r)^2} \left( - 30 \epsilon_H + 18 \eta_H -
24 [\frac{m_P}{\phi} \sqrt{2 \epsilon_H}] \right) 
\,, 
 \label{running}
\eea
with the derivatives of the slow-roll parameters and $r$ given by:
\bea
\eta_H^\prime &=& -\frac{\epsilon_H}{1+r}( \xi_H - 2 \eta_H)\,, \\
\epsilon_H^\prime &=& -\frac{2 \epsilon_H}{1+r}( \eta_H - 2
\epsilon_H) \,, \\
(\frac{m_P}{\phi}\sqrt{2\epsilon_H})^\prime  &=&\left(\frac{m_P}{\phi}\right)\frac{\sqrt{2\epsilon_H}}{1+r} 
\left(\frac{m_P}{\phi}\sqrt{2\epsilon_H} + 2 \epsilon_H - \eta_H
\right) \,,\\
r^\prime &=& \frac{r}{1+7 r} ( 10 \epsilon_H
 +8\frac{m_P}{\phi}\sqrt{2 \epsilon_H} - 6 \eta_H) \,,
\eea
where $\xi_H= 2 m_P^2 V^{\prime \prime \prime}/V^\prime$. 
For the models considered in this section, we have that  
$\epsilon_H = ((\phi/m_P) \eta_H/(n-1))^2/2$, and therefore
typically  $\epsilon_H \ll \eta_H$,  so keeping only the leading 
 terms in  Eq. (\ref{ns}) and (\ref{running}), we end with  
\bea
n_S -1 &\approx& \frac{3 \eta_H }{7 r}\left( \frac{7 - n}{n -1} +
\left(\frac{\phi}{m_P}\right)^2 \frac{3 \eta_H}{2 (n-1)^2}\right)
  \,, \label{nsapprox}
\eea
and
\bea
n_S^\prime & \approx&
-3 \left(\frac{\eta_H}{7r}\right)^2 \left(\frac{n(7-n)}{(n-1)^2}  
+\left(\frac{\phi}{m_P}\right)^2 \frac{(14+10 n - 17 n^2)}{(n-1)^4} \eta_H\right)
\label{runapprox} \,.\\
\eea  
Notice that the spectral index if of the order of 
$O(\eta_H/r)$, whilst the running is of the order of
 $O((\eta_H/r)^2)$. 
Therefore, the same condition needed to have 
slow-roll in the strong dissipative regime will avoid having a too
large spectral index. We will have a  blue-tilted 
spectrum when $n \leq 7$, including the case of $n=0$, i.e., the
logarithmic potential. This is disfavoured by the data \cite{WMAP} when there  
is no running of the spectral index and no tensor contribution to 
the spectrum, otherwise the data is not conclusive. For example with 
non-negligible running, we have the allowed range $0.97 < n_S < 
1.21$ \cite{kinney}, with the running in the range $-0.13 < d n_S/d
  \ln k < 0.007$. The more negative running, the more blue-tilted the
 spectrum can be, which would be the case for $ 0 < n < 7$ with
 $n_S^\prime  \approx -(n_S-1)^2 n/(3(7-n))$.  

Given that the
 field decreases during inflation, so does
 $\eta_H/r$, and also $\rho_R/V$ (or equivalently $T/H$) for any power
 $n \neq 0$, being constant for the
 logarithmic potential $n=0$. Therefore, the energy density in
 radiation will never dominate in this regime. 
 On the other hand, $\phi/T$ diminishes for $n < 4$, but $r$  only for
 $n > 2$. Therefore, for a logarithmic or 
 quadratic potential once the system is brought into the strong
 dissipative regime stays there until the end. Indeed we may start in
 the weak dissipative regime, with $T>H$ but $r< 1$, and it will evolve
 into $r>1$. In the weak dissipative regime the inflaton evolution is
 given by the standard slow-roll equation, but the $T$ behaves like:
\be
\frac{T}{H} \simeq \frac{C_\phi a^4}{36 C_R}
 \left(\frac{\phi}{H}\right)^{2(n-2)} \,,
\label{thw}
\ee
and then
\be
r\simeq \frac{C_\phi}{3} \left(\frac{C_\phi a^4}{36 C_R}\right)^3 
 \left(\frac{\phi}{H}\right)^{2(3n-7)} \label{rw}\,.
\ee
Therefore, for $n=0$ both  $T$ and $r$ grow until $T$ reaches a constant value
when $r \geq 1$; and we may have the interesting situation of a transition
from ``cold'' $\rightarrow$ ``weak warm'' $\rightarrow$
``strong warm'' inflation. For $n=2$,  $T$ remains constant until it
diminishes in the strong dissipative regime, so that the parameter
space divides into either cold or warm inflation, but the system can
evolve from weak to strong dissipation. In the weak dissipative
regime,  quantum fluctuations of the inflaton field will also have a
thermal origin, with an amplitude \cite{Moss:wn,Berera:1995wh}:
\be
P_{\cal R}^{1/2} \simeq \left|\frac{3 H^3}{V_\phi}\right| \sqrt{\frac{T}
  {H}} \simeq 3 \left(\frac{H}{\phi(0)}\right) \sqrt{\frac{C_\phi}
  {36 C_R}}\,, \label{prw} 
\ee
where in the second equality we have used Eq. (\ref{thw}); the
spectral index in given by:
\be
n_S \simeq 1 - 2 \eta_H + 2 \epsilon_H \,,
\ee
Notice that even in the weak dissipative regime  a logarithmic
potential gives rise to blue-tilted spectrum, while for $n >0$ we would
have a red-tilted spectrum, just the reverse than the standard cold
predictions.

For larger powers $n>2$, we  have the opposite behavior, and even if
 inflation starts in the strong  dissipative regime it will evolve
 towards the weak and the cold  regime. When this happens before the
 last 50-efolds of inflation,  then dissipation becomes irrelevant. 

In the following we will briefly consider the cases $n=0,2,4$
 separately.  
 The same than for the monomial
 potentials, we are  interested in getting the constraints on the
 parameter values  ${\cal N}$ and $g_*$, which will depend on the power
 $n$, by demanding first having  50 e-folds of inflation with $r\geq
 1$, and then getting the right primordial spectrum. 

\vspace{0.2cm}

{\bf Case $n=0$: Hybrid logarithmic potential.}  For the potential
Eq. (\ref{vshybrid}) we have that the field value decreases
exponentially as:
\be
\phi \simeq \phi(0) exp[(\eta_H/r) N_e] \,, \label{phil}
\ee
where the ratio $\eta_H/r$ is  approximately constant, and given by:
\bea
\frac{\eta_H}{r} &\simeq& -\left(\frac{64 a^2 C_R^3}{C_\phi^4}\right)^{1/7} \,,
\label{thl}
\eea
with $a^2=\beta V_0/H^4$. The temperature as mentioned before is
approximately constant and given by Eq. (\ref{THhybrid}) with
$n=0$, while the ratios $\phi/T$ and $r$ decreases/increases
respectively like: 
\bea
\frac{\phi}{T} &\simeq& \left(\frac{\phi(0)}{H}\right)\left(\frac{4
  C_\phi C_R}{a^4}\right)^{1/7} exp[(\eta_H/r)N_e] \,, \\
r &\simeq& \frac{1}{3}\left(\frac{H}{\phi(0)}\right)^2\left(\frac{ a^4 
  C_\phi^{4/3}}{4 C_R}\right)^{3/7} exp[-(\eta_H/r)N_e] \,
\label{rl}
\eea
where we have used Eqs. (\ref{phiThybrid}), (\ref{rhybrid}) together
with (\ref{phil}).  
The lower bound on ${\cal N}$ is obtained in the limiting case for slow-roll
warm inflation, when $T/H \simeq 1$, $r(0) \simeq 1$ and
$(\phi/T)_{N} \simeq 10$, where the subscript ``N'' means $N_e$ e-folds
after, which gives:
\be
{\cal N} \simeq 11.87 exp[-0.23 N_e g_*^{1/2}/{\cal N}^{1/2}] \,.
\ee
For example, with $g_*\simeq 10$,
$N_e=50$  we have the lower bound: ${\cal  
  N}\simeq 180$, $\phi(0)/H \simeq 150$, $a^2 \simeq 244$,
$\eta_H/r\simeq -0.054$; with $g_*\simeq 228.75$, $N_e=50$  we have: 
${\cal N}\simeq 1350$, $\phi(0)/H \simeq 1.2 \times 10^3$, $a^2 \simeq 5.2
\times 10^3$, and $\eta_H/r\simeq -0.1$. Those values of 
${\cal N}=N_\chi N_{decay}^2$ are quite in the range of a
realistic model; for example with $\chi$ in the {\bf 126} or {\bf 351}
of $SO(10)$ ($E_6$), and $N_{decay}^2 \approx O(10)$, one can
expect having ${\cal N}$ in the range of a few thousands. However,
for such values we always get a too large 
amplitude of the primordial spectrum, Eq. (\ref{pr}), which using
Eqs. (\ref{thl}) and (\ref{rl}) can be written as: 
\be 
P_{\cal R}^{1/2} \simeq 0.4 \left( \frac{H}{\phi(0)}\right)^{3/2}
\left( \frac{C_\phi^{18} a^6}{C_R^{17}}\right)^{1/28} \,.
\ee
Therefore, in order to match the amplitude of the primordial spectrum
with WMAP's value we need a larger initial value of the field
$\phi(0)$, but then  the value of $C_\phi$ (${\cal N}$) has to be
larger in order to stay  within the low $T$ approximation, with
$\phi/T \geq 10$; at the same time we  
need to increase $a^2$ in order to keep $r(0) \geq 1$. Satisfying
WMAP's constraints requires then ${\cal N} \gtrsim 5\times 10^5$, rather
large, with $\phi(0)/H \gtrsim 2 \times 10^{6}$ and $a^2 \gtrsim 8\times
10^{11}$. 
 
Having inflation in the weak warm regime, and  a transition from weak
to strong dissipation at the end,  might in principle help in fixing
the amplitude of the spectrum to lower values. We still need to  
impose that (a) $T/H> 1$, Eq. (\ref{thw}), (b) we get enough
inflation, i.e., $N_e \approx 50$. These translates into ${\cal N}
\geq 0.05 g_*/\eta_H^2$, with $\eta_H\leq 1/(2 N_e)$, and therefore we
have the lower bound ${\cal N} \gtrsim 0.2g_*N_e^2$. For example for
$g_*= 288.75$ and $N_e \simeq 50$ we have  ${\cal N} \gtrsim 1.2\times
10^5$, which again is rather large. Going from weak to strong
dissipation earlier and 
having only 10 e-folds in the weak regime in principle is a
possibility for smaller values of ${\cal N}$, but then again we cannot
keep the low $T$ approximation  and $\phi/T \geq 10$ for the remaining
e-folds of inflation. However one has to
bear in mind that we are not properly taking into account how
inflation ends, which requires going beyond the  analytical
approximations. The transition from the low to the high $T$ is not
necessarily  instantaneous as we have implicitly  assumed in this
paper, and may last another few  e-folds. Counting all together can
bring again the values of the parameters into realistic values.   

\vspace{0.2cm}

\begin{figure}[t] 
\hfil\scalebox{0.5} {\includegraphics{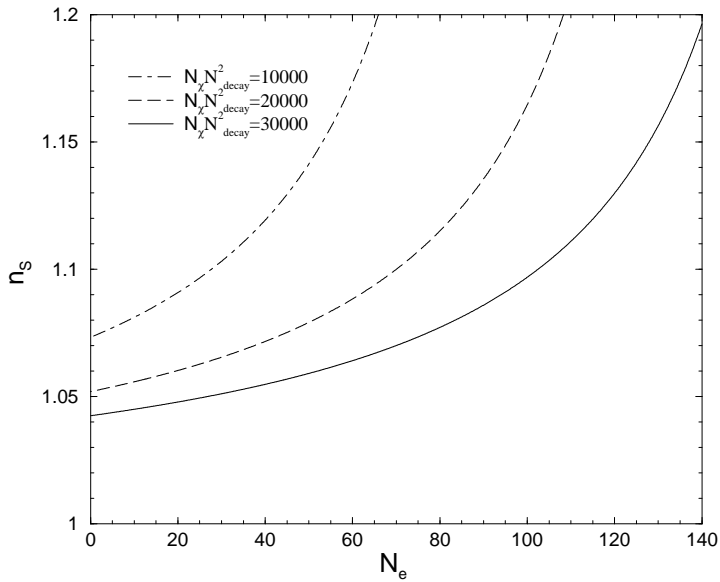}}\hfil
\caption{Hybrid quadratic potential: Spectral index depending on the
  no. of e-folds to the end of inflation, for different values of
  ${\cal N}$. We have taken: $g_*=230$, $\phi(0)/m_P=0.21$, $\eta_H
  =3$, and $V_0^{1/4}/m_P=3\times 10^{-4}$.}
\label{plot2}
\end{figure}

{\bf Case $n=2$: Hybrid Quadratic potential.} The main restriction for
this model 
comes from  keeping the ratios $T/H$ and $\phi/T$ within the range of the
low $T$ approximations. During slow-roll the dissipative ratio $r$
increases, so its initial value $r_0$ gives the lower bound on this
parameter. On the other hand, we still require $(\eta_H/r)_0 = a^2/(3
r_0) < 1$ in order to have slow-roll warm inflation. By using
Eqs. (\ref{phihybrid}), and (\ref{rhybrid}), we can rewrite
Eqs. (\ref{THhybrid}), (\ref{phiThybrid}), in terms of $r_0$ and
$\eta_H/r_0$ like:
\bea
\frac{T}{H} &\simeq& \left(\frac{C_\phi}{C_R}
\right)\left(\frac{\eta_H}{2 r_0} ( 1 + \frac{2 \eta_H}{7 
  r_0} N_e)^{-1} \right)^{2} \,, \\ 
\frac{\phi}{T} &\simeq& \left(\frac{C_\phi}{C_R^{1/2}}\right)
\left(\frac{3 \eta_H}{2 r_0}\right) \left( 3 r_0 ( 1 + \frac{2 \eta_H}{7 
  r_0} N_e)^{3} \right)^{-1/2} \,, 
\eea
Thus, unless $C_R$ is too small, the condition $\phi/T>10$ will be
fulfilled  if we have $T>H$. This in turn translates into a lower bound
for ${\cal N}$: 
\be
{\cal N} \gtrsim 0.052 g_* \left( \frac{r_0}{\eta_H} + \frac {2 N_e}{7} \right)^2 \,. \label{Ngstarqd}
\ee
From the approximated expression for the spectral index,
Eq. (\ref{nsapprox}), if we want to keep $n_S$ within the observable
range, we require $\eta_H/r_0 \leq 0.093$, which for $N_e \simeq 50$ gives the
lower bound ${\cal N} \gtrsim 32.5 g_*$. Again, having slow-roll warm
inflation for example with $g_*\simeq 228.75$ needs ${\cal N} \gtrsim
7500$, but for $g_* \simeq 10$ it only requires ${\cal N} \gtrsim
325$. As an 
example, in fig. (\ref{plot2}) we have plotted the predicted spectral
index depending on the no. of e-folds left to the end of inflation, for
${\cal N} \equiv N_{\chi} N_{decay}^2 = 10000, 20000, 30000$ 
and $g_*= 228.75$. The
corresponding spectral index 
of the primordial spectrum would be that at around 50-55 e-folds,
which is always $n_S < 1.2$. The value of the running can be obtained
from Eq. (\ref{runapprox}), and it is given respectively by
$n^\prime_S \simeq -2.5\times 10^{-3},\; -8.5 \times 10^{-4},\; -4.7
\times 10^{-4}$.    

We can estimate the amplitude of the primordial spectrum,
Eq. (\ref{pr}) in terms of the parameters $r_0$ and $\eta_H/r_0$: 
\be
P_{\cal R}^{1/2} \simeq 4 \sqrt{3}\left(\frac{4 \pi r_0^3}{3}\right)^{1/4}
 \left(\frac{r_0}{\eta_H} \right)^3
\frac{C_R^{1/2}}{C_\phi^{3/2}} \,.
\ee
Having a not too large amplitude for the primordial spectrum
prefers values of $\eta_H/r_0$ as large as possible, but not too large
values of 
$r_0$. For values of ${\cal N}$, $g_*$ within the range of
Eq. (\ref{Ngstarqd}), the amplitude remains below say\footnote{Given
  the uncertainties in the analytical estimations, and in
  Eq. (\ref{pr}), it would not make sense to impose the specific WMAP
  value as a constraint, but an order of magnitude estimation would be
  sufficient. Furthermore, we have checked numerically that when the
  primordial spectrum is originated with $r_0$ not much larger than
  unity, the value of the amplitude tends to be lower than that given
  in Eq. (\ref{pr}).}  $10^{-4}$ for values of $r_0$ of the order  of
$O(10)$.  Therefore in this kind of models parameter values can be
found giving rise to the right order of magnitude for the primordial
spectrum in the strong dissipative regime, but the stronger constraint
comes from getting a not too blue-tilted spectrum.  

\vspace{0.2cm}

{\bf Case $n \geq 4$: Hybrid quartic and higher powers.} 
In this case we have that $T/H$ and $r$, Eqs. (\ref{THhybrid}) and
(\ref{rhybrid}),  both decrease during
inflation, while on the contrary $\phi/T$ increases. Therefore, it is
not difficult to remain within the range of the low $T$ 
approximation, but on the other hand strong dissipation with $r \geq
1$ will  last only 
a few e-folds.  With $n=4$, having 50 e-folds in the strong
regime requires for example ${\cal N}=N_\chi N_{decay}^2 \gtrsim
10^3$ for $g_* \simeq 10$, and ${\cal N} \gtrsim
10^4$ for $g_* \simeq 228.75$. In addition if we want to get the right
amplitude for the 
spectrum Eq. (\ref{pr}), and spectral index, this value increases by
one order of magnitude, as  we need to adjust the values of the quartic
coupling $a^2=\lambda$ to rather small values, and then the initial value of
the field to larger values to have $\phi(0)/T \geq 10$. Numbers do not
change much if we demand 10 or 50 e-folds of inflation in the strong
dissipative regime.  

\vspace{0.2cm}

{\bf Comments: Ending inflation} 

We have seen that depending on the values of the field
($\phi(0)$) and couplings ($V_0$, $\beta$) in the inflationary
potential with $n=0,2$, we may have either weak or strong dissipative
regime during $N_e$ e-folds, but
for rather large values of the parameter ${\cal N}=N_\chi
N_{decay}^2$. However, those large values are not required  
for having  inflation in the weak/strong dissipative regime, but in 
order to match the predicted values of the spectrum with the cosmological
observations, including having enough inflation in the low $T$ regime.
With lower values of
the multiplicity parameter, in the range of a few thousands or so,
dissipative effects are relevant for the inflationary dynamics, and
indeed the tendency is to bring the system into the strong dissipative
regime, first in the low $T$ regime, and later  into the
high $T$ regime, where the analytical approximations break down. From
this point of view, cosmological observations rule out large regions
of the parameter space in this kind of models. Nevertheless, a
quantitative statement on the values of the parameters needs in turn
to properly take into account the dynamics at the end of
inflation. There are some important  corrections that have to be
included before ruling out any of the models presented here because 
they do not fit cosmological observations. First of all,
as already mentioned, how the system interpolates between the low and the
high $T$ regime, and for how long (how many e-folds) this period
lasts. In addition, in hybrid models as the inflaton evolves towards
the critical value $\phi_c$ the mass of the waterfall field decreases
accordingly, approaching the tachyonic instability, with $m_\chi^2 = 2
g^2 (\phi^2-\phi_c^2)$ . The dissipative coefficient
Eq. (\ref{upsilon0}) depends on the ratio of $T^3/m^2_\chi$, and thus 
close to the critical value it will get a rather large enhancement
factor, with    
\be
 \Upsilon_\phi \simeq C\phi \left(\frac{\phi^2}{\phi^2 - \phi_c^2}\right)^3
\frac{T^3}{\phi^2} \,.   
\label{upshybrid}
\ee
Although it may look at a first glance that this enhancement of
$\Upsilon_\phi$, and therefore $r$, will bring the system faster into
the high $T$ regime, the increase of the extra friction in the inflaton
evolution  tends to slow down the field, and as a consequence also the
evolution of the ratios $T/H$ and $\phi/T$. This again provides some
extra e-folds of inflation in the warm dissipative regime. Finally, we
will have to take into account thermal corrections in the scalar
masses, mainly $m_\chi$, which can be non negligible towards the end
of inflation in the high $T$ regime. Nevertheless, in order to take
into account these effects one has to resort to numerical calculation,
which are beyond the scope of this paper \cite{wprogress}. The results
presented here  can be seen  as a kind of worst case scenario, and the
bounds on the parameters will be relaxed once we give up the
requirement of having the full 50-60 last e-folds in the low $T$ range.

\section{Summary}
\label{sect3}

This paper has made an important step in understanding
dissipative effects and particle production during
the inflationary expansion period.
These results are of general significance in understanding
inflationary dynamics, since they break preconceived ideas
set early in the development of inflation that the dynamics
during inflation has negligible dissipative and particle production
effects.  The results in this paper show that picture does not
express the general case and that dissipative effects become important
in very generic models.
Moreover as discussed in this paper and 
elsewhere \cite{warm,Berera:1999ws,bb1,bgr,br,ian,arjunspectrum,hmb}, the
consequences of dissipation during inflation are
nontrivial with respect to observable signatures and
model building prospects.

The technical problem of determining dissipative effects
during inflation should not be underestimated.
To appreciate this point,
it is useful to compare this
warm inflation problem to the standard cold inflation problem.
In the latter, the particle production phase is pictured as
entirely separated from the inflationary expansion phase;
this problem has been intensely studied for well over
two decades and is still not fully understood.
The warm inflation problem is technically much more difficult,
since particle production and inflationary expansion
are meant to be occurring concurrently
and this problem has been examined for
a much shorter amount of time.

The basic problem of determining dissipative effects during
inflation can be posed as follows.
A background inflaton field is evolving slowly
in time as the Universe inflates.  This field is coupled to other
fields, thus in general dissipative and particle production
effects can be expected.  This is a nonequilibrium
situation, with the central problem being to
determine what is the statistical state of the Universe.
To address this problem from first principles quantum
field theory has required building up the necessary
knowledge through an interplay of investigating nonequilbrium
approximations and then testing their relevance to the actual inflation
problem.
Ideally one could imagine trying to solve this problem
through numerical calculations and computer simulations,
but the problem is too complex to immediately do a very general
treatment this way.  Ultimately the goal is to achieve
some sort of reliable treatment of the problem by these
methods.  However before that can be achieved, a clearer idea
of the approximate state of the system is needed.
This is the direction that has been developing in the past
few years \cite{bgr}.  The one baseline one has to go on
is the assumption the statistical state remains close to
thermal equilibrium throughout evolution.
There are obvious deviations from equilibrium due to
the evolution of the background inflaton field
and due to production of light particles
and their rescattering with the background field.
Nevertheless, thermal equilibrium is a good limiting case to examine
since reliable and unambiguous calculations can be made
of dissipation and particle production.
However the assumption of equilibrium is a conservative one,
from which one anticipates
dissipative effects will be minimal.
As such this approximation is vitally important in
determining a minimal level of warm inflation.

With these considerations in mind, the results in this paper are
all the more encouraging. We have shown that all the simple
inflation models, those with monomial and hybrid potentials,
have warm inflation regimes.  Moreover the two distinguishing
model building features of warm inflation, inflaton mass
bigger than the Hubble parameter so complete avoidance to
the ``eta problem'' and field amplitudes below the Planck
scale even in monomial potentials, have both been
verified explicitly in models.
Although improvements on the equilibrium approximation to
obtain the correct nonequilibrium state will quantitatively change
the results from those in this paper, these qualitative features
will persist and so have been firmly established as realizable from
realistic quantum field theory models.

The quantitative changes from improving on the equilibrium
approximation should lead to a decrease in the total number of fields
and so to even more simpler models that yield warm inflation.
In this paper, within the equilibrium approximation, we found
models realizing warm inflation requiring a minimum of
$N_\chi \approx O(10^3/N^2_{decay})$ fields.  This is
within the realm of realistic model building with particle physics
GUT models such as $SO(10)$ or $E_6$, with the $\chi$ field in the
{\bf 210} or {\bf 351}, and a decay factor $N_{decay}\sim
O(3-4)$.  
To gain an idea of how much these models can be improved
from the nonequilibrium treatment can be gained by examining
the results already studied for warm inflation using
the quasiparticle approximation \cite{br,br05,Hall:2004zr,bb1}.
This approximation has been studied in the context of
quantum field theory and cosmology applications since
the work of Morikawa and Sasaki in the mid 1980s \cite{Morikawa:dz}.
It is motivated by generic features known of many-body systems
in condensed matter physics.  No derivation of this approximation
has been made in quantum field theory, but there is
some suggestive numerical evidence in support of it \cite{aartsberges}.
One of the next steps in the development of warm inflation dynamics
is to understand by how much the equilibrium approximation deviates
and in particular how closely it tends to the quasiparticle
approximation.

In summary, this paper has accomplished two things.  
On the side of principle,
due to the thermal equilibrium approximation underlying
all the results, this paper has calculated minimal
expectations of radiation production and dissipative effects
during inflation.  On the side of application, this paper has
identified and developed a reliable methodology for
performing warm inflation calculations.  Although the
results found here are only on the periphery of
usefulness for model building, the methodology developed in
this paper gives a foothold from which further improvements
can be made.  Aside from the major modifications already mentioned
of expanding the equilibrium approximation to a more accurate
nonequilibrium treatment, there are also several improves within
the equilibrium approximation that can be made, and
would lower the number of fields necessary, thus increase
the scope of model building prospects.
One modification is to move away from the strict low temperature regime
into the intermediate temperature regime.  This is technically
much more difficult since now all finite temperature effective
potential corrections will have to be accounted for.
However the results of this paper already anticipate
that the relaxation of this condition will increase
dissipative effects, lower the total number of fields needed
and help resolve issues of exiting the inflation epoch.
A second improvement is treating higher order calculations of
the dissipation coefficient.
This dissipative coefficient is closely related to the shear viscosity,
as observed in \cite{bgr}, for which resummation calculations \cite{jeon}
have shown provide as much as a factor four enhancement.
One anticipates similar enhancements could occur also
for the dissipative coefficients.

\begin{acknowledgments}
The authors thank Ian Moss for valuable discussions, and Lisa Hall for
comments. 
AB was funded by the United Kingdom Particle Physics and
Astronomy Research Council (PPARC).

\end{acknowledgments}

\end{document}